\documentclass[english]{revtex4}
\usepackage[T1]{fontenc}
\usepackage[latin9]{inputenc}
\usepackage{amsmath}

\makeatletter
\usepackage{babel}
\makeatother

\begin{document}

\title{Stringy Generalization of the First Law of Thermodynamics for Rotating
BTZ Black Hole with a Cosmological Constant as State Parameter}

\author{Eduard Alexis Larrañaga Rubio}

\email{eduardalexis@gmail.com}

\affiliation{National University of Colombia}

\affiliation{National Astronomical Observatory (OAN)}

\begin{abstract}
In this paper we will show that using the cosmological constant as
a new thermodynamical state variable, the differential and integral
mass formulas of the first law of thermodynamics for asymptotic flat
spacetimes can be extended to be used at the two horizons of the (2+1)
dimensional BTZ black hole. We also extend this equations to the stringy
description of the BTZ black hole, in which two new systems that resemble
the right and left modes of effective string theory, are defined in
terms of the inner and outer horizons.
\end{abstract}
\maketitle

\section{Introduction}

Bekenstein and Hawking showed that black holes have non-zero entropy
and that they emit a thermal radiation that is proportional to its
surface gravity at the horizon. These two quantities are related with
the mass through the identity

\begin{equation}
dM=TdS,\end{equation}

that is called \emph{the first law of balck hole thermodynamics}\citet{hawking,bekenstein}.But
when the black hole has other properties as angular momentum $\mathbf{J}$
and electric charge $Q$, the first law is generalized to 

\begin{equation}
dM=TdS+\Omega dJ+\Phi dQ,\end{equation}

where $\Omega=\frac{\partial M}{\partial J}$ is the angular velocity
and $\Phi=\frac{\partial M}{\partial Q}$ is the electric potential.
The corresponding integral Bekenstein-Smarr mass formula, and is given
by

\begin{equation}
M=\Phi Q+\frac{\left(D-2\right)}{\left(D-3\right)}\Omega J+\frac{\left(D-2\right)}{\left(D-3\right)}TS.\label{eq:bekensteinsmarr}\end{equation}

Gauntlett et. al.\citet{gauntlett} have proved that both, the differential
and integral expressions, hold for asymptotically flat spacetimes
with any dimension $D\geq4$ . For rotating black holes in anti-deSitter
spaces, Gibbons et. al.\citet{gibbons} have shown that the differential
expression hold for $D\geq4$, but the integral expresion is not satisfied.
To rectify this situation Caldarelli et. al. \citet{caldarelli} use
a cosmological constant considered as a new thermodynamical variable,
and recently, Wang et. al. \citet{wang} use this idea to show that
differential and integral expressions are valid also in $\left(2+1\right)$
dimensions for BTZ black holes with angular momentum and Kerr-de Sitter
spacetimes.

In this paper we will consider the BTZ black hole in $\left(2+1\right)$
dimensions to show that considering a cosmological constant as a thermodynamical
state variable, both, the differential and integral mass formulas
of the first law, can be extended to be used on the two horizons of
the metric. The same procedure is also applicable in the stringy description
of the BTZ black hole, in which two new systems that resemble the
right and left modes of effective string theory, are defined in terms
of the inner and outer horizons. Here we obtain the generalized force
conjugate to the cosmological constant term in all the cases.

\section{The Rotating BTZ Black Hole}

The rotating BTZ black hole \citet{martinez} is a solution of $\left(2+1\right)$
dimensional gravity with a negative cosmological constant $\Lambda=-\frac{1}{l^{2}}$.
Its line element can be written as

\begin{equation}
ds^{2}=-\Delta dt^{2}+\frac{dr^{2}}{\Delta}+r^{2}d\varphi^{2},\end{equation}

where the lapse function is

\begin{equation}
\Delta=-M+\frac{r^{2}}{l^{2}}+\frac{J^{2}}{4r^{2}}.\end{equation}

This solution has two horizons given by the condition $\Delta=0$,

\begin{equation}
r_{\pm}^{2}=\frac{Ml^{2}}{2}\left[1\pm\sqrt{1-\frac{J^{2}}{M^{2}l^{2}}}\right].\label{eq:radii}\end{equation}
The mass $M$ and angular momentum $J$ of the black hole can be written
in terms of this horizons as

\begin{equation}
M=\frac{r_{+}^{2}+r_{-}^{2}}{l^{2}},\label{eq:mass1}\end{equation}

\begin{equation}
J=\frac{2r_{+}r_{-}}{l}.\end{equation}

The Bekenstein-Hawking entropy associated with the black hole is twice
the perimeter of the outer horizon,

\begin{equation}
S=4\pi r_{+},\end{equation}
and therefore, the mass can be written as

\begin{equation}
M=\frac{r_{+}^{2}}{l^{2}}+\frac{J^{2}}{4r_{+}^{2}}=\frac{S^{2}}{16\pi^{2}l^{2}}+\frac{4\pi^{2}J^{2}}{S^{2}}.\label{eq:massformula1}\end{equation}

This expression can be re-written as \citet{wang}

\begin{equation}
M=\frac{1}{2}TS+\Omega J,\label{eq:generalIntFirstLaw}\end{equation}

As is well known, we can associate a thermodynamics to the outer horizon
when treated as a single thermodynamical system. With the four laws
associated with this horizon one can describe the Hawking radiation
process. On the other hand, it is also possible to consider the inner
horizon as an independient thermodynamical system and associate it
a set of four laws that are related with the Hawking absorption process
\citet{wu2}. Therefore, the integral and differential mass formulae
can be written for the two horizons,

\begin{eqnarray}
M & = & \frac{r{}_{\pm}^{2}}{l^{2}}+\frac{J^{2}}{4r{}_{\pm}^{2}}\label{eq:massformula}\end{eqnarray}
If we consider an invariable $l$, this expression let us calculate
the surface gravity and angular velocity for the black hole as

\begin{eqnarray}
\kappa_{\pm} & = & \frac{1}{2}\left.\frac{\partial M}{\partial r_{\pm}}\right|_{J}=\frac{r_{\pm}}{l^{2}}-\frac{J^{2}}{4r_{\pm}^{3}}\label{eq:gravity}\\
\Omega_{\pm} & = & \left.\frac{\partial M}{\partial J}\right|_{r_{\pm}}=\frac{J}{2r_{\pm}^{2}}=\frac{r_{\mp}}{r_{\pm}l},\label{eq:angularvelocity}\end{eqnarray}

while the entropy and temperature associated with each horizon are

\begin{eqnarray}
S_{\pm} & = & 4\pi r_{\pm}\label{eq:entropy}\\
T_{\pm} & = & \frac{\kappa_{\pm}}{2\pi}.\label{eq:temperature}\end{eqnarray}

If we make the variation of the mass formula (\ref{eq:massformula})
with respect to $S$ and $J$, we obtain the first law of black hole
thermodynamics in differential form for each horizon, 

\begin{equation}
dM=TdS+\Omega dJ.\end{equation}
But if we use equations (\ref{eq:temperature}) and (\ref{eq:angularvelocity}),
the mass formula can be written as

\begin{equation}
M=\frac{1}{2}T_{\pm}S_{\pm}+\Omega_{\pm}J,\end{equation}
that does not correspond to the Bekenstein-Smarr formula (\ref{eq:bekensteinsmarr})
with $D=3$ for any horizon. 

Note that the product $T_{\pm}S_{\pm}$ in this case is give by 

\begin{equation}
T_{\pm}S_{\pm}=\frac{2r_{\pm}^{2}}{l^{2}}-\frac{J^{2}}{2r_{\pm}^{2}}=\frac{2r_{\pm}^{2}}{l^{2}}-\frac{2r_{\mp}^{2}}{l^{2}},\end{equation}
while the product $\Omega_{\pm}J$ is

\begin{equation}
\Omega_{\pm}J=\frac{r_{\mp}^{2}J}{r_{\pm}l}=\frac{2r_{\mp}^{2}}{l^{2}}.\end{equation}
Then, we have that

\begin{equation}
T_{\pm}S_{\pm}+\Omega_{\pm}J=\frac{2r_{\pm}^{2}}{l^{2}},\end{equation}
cannot fit into the Bekenstein-Smarr formula.

\section{Cosmological Constant as a New Thermodynamical State Variable}

Now, we will try to rectify this situation by considering the effect
of a varying cosmological constant. When considering the cosmological
constant as a new state variable, the first law in differential and
integral forms for each horizon must be corrected to be

\begin{eqnarray}
dM & = & T_{\pm}dS_{\pm}+\Omega_{\pm}dJ+\Theta_{\pm}dl\\
0 & = & T_{\pm}S_{\pm}+\Omega_{\pm}J+\Theta_{\pm}l,\end{eqnarray}

where $\Theta_{\pm}$ are the generalized forces conjugate to the
parameter $l$ on each horizon. These generalized forces are given
by

\begin{eqnarray}
\Theta_{\pm} & = & \left.\frac{\partial M}{\partial l}\right|_{S_{\pm},J_{\pm}}=-\frac{2r_{\pm}^{2}}{l^{3}},\end{eqnarray}

All the thermodynamical quantities, including the generalized force
for the outer horizon in last expression coincide with the reported
results \citet{wang}. Now we will turn our attention to the stringy
analogy of the thermodynamics of the BTZ black hole to see how can
we generalize the Bekenstein-Smarr formula in that case.

\section{Stringy Thermodynamics for the BTZ Black Hole}

The inner and outer horizons of the BTZ black hole let us define two
new thermodynamical systems that resemble the right and left modes
of string theory. Using the notation of \citet{larr2}, the R and
L systems are defined by

\begin{equation}
\mathcal{P}_{R,L}=r_{+}\pm r_{-},\end{equation}

where $\mathcal{P}$ is the ''reduced'' perimeter. For the R-system,
the mass can be written as

\begin{equation}
M=\frac{\mathcal{P}_{R}^{2}}{l^{2}}-\frac{J}{l}.\end{equation}

Following the above treatment, the generalized force associated with
the R-system is obtained as

\begin{equation}
\Theta_{R}=\left.\frac{\partial M}{\partial l}\right|_{\mathcal{P}_{R},J}=-\frac{2\mathcal{P}_{R}^{2}}{l^{3}}+\frac{J}{l^{2}},\end{equation}

and can be written in terms of $\Theta_{\pm}$ as

\begin{equation}
\Theta_{R}=\Theta_{+}+\Theta_{-}+\frac{\Omega_{R}J}{l},\end{equation}

where $\Omega_{R}=-\frac{1}{l}$, as reported in \citet{larr2}. 

On the other hand, for the L-system, the mass is given by

\begin{equation}
M=\frac{\mathcal{P}_{L}^{2}}{l^{2}}+\frac{J}{l}.\end{equation}

Thus, the generalized force associated with the L-system is

\begin{equation}
\Theta_{L}=\left.\frac{\partial M}{\partial l}\right|_{\mathcal{P}_{L},J}=-\frac{2\mathcal{P}_{L}^{2}}{l^{3}}-\frac{J}{l^{2}},\end{equation}

or, in terms of $\Theta_{\pm}$, as

\begin{equation}
\Theta_{L}=\Theta_{+}+\Theta_{-}+\frac{\Omega_{L}J}{l},\end{equation}

where we have used $\Omega_{L}=\frac{1}{l}$. Therefore, the generalized
force associated with the R and L systems are given by the relation

\begin{equation}
\Theta_{R,L}=\Theta_{+}+\Theta_{-}+\frac{\Omega_{R,L}J}{l},\end{equation}

and the corresponding equations for differential and integral first
law are generalized to

\begin{equation}
\begin{cases}
dM & =T_{R,L}dS_{R,L}+\Omega_{R,L}dJ+\Theta_{R,L}dl\\
0 & =T_{R,L}S_{R,L}+\Omega_{R,L}J+\Theta_{R,L}l.\end{cases}\end{equation}

\section{Conclusion}

By considering the cosmological constant as a new thermodynamical
state variable, we have generalized the integral Bekenstein-Smarr
mass formula and differential first law for the two horizons of the
$\left(2+1\right)$ dimensional BTZ black hole. We also show how the
same definition can be used to describe the R and L thermodynamical
systems, constructed from the inner and outer horizons, and that resemble
the right and left modes of string theory. The generalized forces
associated with the cosmological term are also obtained, showing a
great symmetry for the R and L systems. It is interesting to note
that the obtained relations can also be extended to higher dimensions,
as noted in \citet{wang}.

\end{document}